\documentstyle[prl,aps,preprint,epsfig,amsmath]{revtex}

\newif\ifcolor
\colortrue 

\colorfalse 

\begin{document}
\normalsize
\draft
\widetext

\title{Analytical solution of the optimal laser control problem in two-level systems}
\author{Martin E. Garcia}
\address{Departament de F\'{\i}sica, Universitat de les Illes
Balears, E-07071 Palma de Mallorca, Spain,}
\author{Ilia Grigorenko}
\address{Institut f{\"u}r Theoretische Physik der Freien
Universit{\"a}t Berlin, Arnimallee 14, 14195 Berlin, Germany.}
\date{\today}
\maketitle

\begin{abstract}
The optimal control of two-level systems by time-dependent laser
 fields  is studied using a variational theory.
 We obtain, for the first time,  general analytical expressions
 for the optimal pulse shapes leading to
 global maximization or minimization of different physical quantities.
We present solutions which reproduce and improve previous
 numerical results.

\end{abstract}


Optimal laser control in quantum systems is a problem of fundamental
importance for atomic, molecular, solid-state and chemical physics
which has attracted much attention in the last 10 years.  Any
system exhibiting quantum coherence can be subject to optimal control,
which basically consists in the manipulation of the quantum dynamics
by external time-dependent laser fields.  A general procedure for this
manipulation is described by the optimal control theory
(OCT)\cite{rabitz_oct,rabitz_constant}. The time-envelope of the
external field is optimized by experimental or numerical
pulse-shaping techniques in order to drive the wave function $\psi(t)$
of the system to fit a target state $\phi_0$ at a particular control
time $t_{control}$. Thus, the optimal shape $V_{opt}(t)$ of the
external field achieves the maximization of the quantity $|\langle
\phi_0 | \psi(t_{control}) \rangle |$.  Since the pioneering work by
Hudson and Rabitz\cite{hudson}, different approaches to optimal
control have been proposed and experimentally applied in various
contexts\cite{rabitz_oct,rabitz_constant,shapiro,tannor,gerber,woeste,girard,murnane,oliver}. The
major problem of the current theoretical description of optimal
control is that the resulting
equations are of high complexity and must be solved numerically.  Therefore,
  neither the experimental realization nor the theoretical
description of optimal control guarantee that the result of the optimization
 corresponds to the true global extremum of the control problem
considered.  As a functional of the pulse
shape, the physical quantity to be maximized or minimized represents a
hyper surface in a multidimensional space, which might exhibit many
local extrema in which the experimental or numerical procedure can get trapped.

The only way to extract the global extremum from among the multiplicity of
 local extrema is by finding the analytical solution to the control problem.
 In this paper we present for the first time analytical results
for the optimal pulse shapes leading to the global extrema of
 different optimal control problems.

We concentrate on physical situations which can be described by
two-level systems.   Using our approach, we give explicit pulse shapes for
 inducing the maximization of population transfer between two levels
and for the  achievement of self-induced transparency under the
constraint of fixed pulse energy.

 We consider a physical quantity $Q$ which is only nonzero when the system is in
 the excited state, and which we wish to  maximized or minimized within the
 control interval $[T_0,T_1]$. For this purpose an external field of optimal shape
has to be applied.  $Q$ can be seen as a fitness function. We
 write it as a time average, over the time interval $[T_0,T_1]$, of the form
 \begin{equation}
Q = \int_{T_0}^{T_1} {\cal{Q}}(t) dt.
 \end{equation}
 ${\cal{Q}}(t)$ is a fitness density, which is a functional of the
 density matrix $\rho(t)$ of the system and, consequently, also of the
 external field $V(t) \cos (\omega t)$.

 Our approach to derive an equation for the optimal field shape
is based on a Lagrangian  of the form\cite{GGB2002}
\begin{eqnarray}
\label{lagrang_main}
L&=&\int_{T_0}^{T_1}{{\cal{L}}dt}=\int_{T_0}^{T_1}{\Gamma(t)\Big ( \frac{\partial}{\partial t}+
i \hat{\cal{Z}}(t)\Big ) \rho(t) dt} \, + \,  \lambda \int_{T_0}^{T_1}{{\cal {Q}}[\rho(t)] dt}+\lambda_1 \int_{T_0}^{T_1}{V^2(t) dt},
\end{eqnarray}
where ${\cal{L}}$ is the Lagrangian density, $\lambda$ and $\lambda_1$ are Lagrange multipliers and  $\Gamma(t)$
a Lagrange multiplier density\cite{units}.

The first term in Eq.~(\ref{lagrang_main}) ensures that the density
matrix satisfies the quantum Liouville equation with the corresponding
Liouville operator $\hat{\cal{Z}}(t)$. The second term explicitly
includes the description of the optimal control and refers to a
physical quantity to be optimized during the control time interval.
The third term in Eq.~(\ref{lagrang_main}) represents a constraint on
the total energy $E_0$ of the control field\cite{GGB2002}. The
Lagrangian (\ref{lagrang_main}) is the basis of our control theory,
since it allows the derivation of the equations to be fulfilled by the
control field.

 As mentioned before, one of the purposes of this paper is to find
 analytical solutions for the "standard" optimal control problem. This
 means, we search for the maximization or minimization of ${\cal
 {Q}}(t_{control})$ at a particular control time $t_{control}$. Note
 that we can treat this problem as a particular case of the theory
 presented above. We only need to modify the fitness density ${\cal
 {Q}}$ by introducing a delta function as follows
\begin{eqnarray}
\label{objective}
Q = \int_{T_0}^{T_1}{{\cal {Q}} [\rho(t)] \delta(t-t_{control})} dt={\cal {Q}} [\rho(t_{control})],
\end{eqnarray}
 where $t_{control}$ lies within the time interval $[T_0,T_1]$.

 We now apply the theory described above to
 a two-level quantum system with energy levels $
 \epsilon_1$ and $ \epsilon_2$, interacting with an external control
 field of the form $V(t) \cos (\omega t)$, where $\omega$ is the
 carrier frequency. If the resonant condition $\omega = \epsilon_2-
 \epsilon_1$ and adiabaticity criterion $|\dot{V}\omega|\ll|V|^3$
 \cite{shore} for the control fields are satisfied, one can use the
 Rotating Wave Approximation (RWA) to derive the Liouville equation
 for the density matrix
\begin{eqnarray}
\label{RWA}
i \frac{\partial{\rho_{11}}}{\partial{t}}
&=&\mu V(t)(\tilde{\rho}_{12}-\tilde{\rho}_{21})+i\gamma_1 \rho_{22},  \nonumber\\
i \frac{\partial{\rho_{22}}}{\partial{t}}
&=&\mu V(t)(\tilde{\rho}_{21}-\tilde{\rho}_{12})-i\gamma_1 \rho_{22}, \\
i \frac{\partial{\tilde{\rho}_{12}}}{\partial{t}}
&=&\mu V(t)(\rho_{22}-\rho_{11})-i\gamma_2\tilde{\rho}_{12}, \nonumber
 \end{eqnarray}
where $\mu$ is  the dipole matrix element of the two-level system and $\gamma_1,\gamma_2$ are relaxation and dephasing constants, respectively.
 Here we use the notation
$\tilde{\rho}_{12}=\rho_{12} \exp(i \omega t)$ and
$\tilde{\rho}_{21}=\rho_{21} \exp(-i \omega t)$. $\rho_{11}$ and $\rho_{22}$  correspond to the instantaneous occupation of the ground and excited state, respectively.
 Note, that $\rho_{11} + \rho_{22} = 1$ and $\tilde{\rho}_{21}
=\tilde{\rho}_{12}^*$.
 We set the initial conditions as  $\rho_{11}=1,\rho_{22} = \tilde{\rho}_{12}=
\tilde{\rho}_{21}=0$.

As a first application we address the phenomenon known as  self-induced
transparency (SIT)\cite{john,panzarini}.  The problem consists in
finding a  temporal pulse shape for which a light pulse
entering a material propagates without significant losses.
 This effect has been studied theoretically using different approaches\cite{john,panzarini}. However, it has never been considered so far as an optimal control problem. We show below that SIT can be viewed as the search for
  the optimal pulse shape for which losses during propagation are minimized.

 In order to solve the problem from the perspective of  optimal
 control, one can assume, as usual,
 that materials where SIT occurs are collections of
 inhomogeneously broadened two-level systems\cite{john,panzarini}. If
 we assume, in addition, that the material is optically thin (thus, we neglect
changes of the field along the spatial axis), we can
 describe the phenomenon using the theory presented above and obtain
 analytical results. The losses of such a system are proportional to
 the average occupation of the upper level
\begin{equation}
 \Gamma_{loss} \propto Q_{22} = \int_{T_0}^{T_1}{\rho_{22} (t) dt}.
\end{equation}
  Thus, we search for the
 minimization of the physical quantity
 $Q_{22}$, i.e.,  the integral
 of the occupation of the excited level over the time interval $[T_0,T_1]$.
 Since systems exhibiting
 SIT are characterized by long life-times of the excited levels, we
 assume $\gamma_{1,2} T \ll 1$, where $T$ is a characteristic time
  during which the system is in the exited state. $T$ is proportional
 to the inverse Rabi frequency induced by the control field.
 This assumption  permits
 us to neglect relaxation and dephasing effects within the control
 interval and write the solution for occupation of the upper level
 $\rho_{22}(t)$ as
\begin{equation}
\label{ocup22}
\rho_{22}(t)=\sin^2 \left( \theta(t) \right),
\end{equation}
where the pulse area $\theta$ is defined as
\begin{equation}
\label{pulse_area}
\theta(t) = \mu \int_{T_0}^t dt' V(t').
\end{equation}
Thus, the Lagrangian density is a function of the pulse area and its first derivative ${\cal{L}}(\theta,\dot{\theta})$ and is given by
\begin{equation}
\label{lagrang_simple}
{\cal{L}} = \sin^2 \left( \theta(t) \right)+\lambda  \frac{\dot{\theta}^2(t)}{ \mu^2}.
\end{equation}
In Eq.~(\ref{lagrang_simple}), and also in the rest of the paper, we omit the Liouville term since we include the analytical expression (\ref{ocup22})
for the density matrix which solves  the Liouville equation. The  Euler-Lagrange equation derived from the above Lagrangian density reads
\begin{equation}
\label{sine-gordon}
2\lambda \ddot{\theta}(t)- \mu^2 \sin(2  \theta(t))=0.
\end{equation}
Eq.~(\ref{sine-gordon}) is of the second order and
 requires two boundary conditions.
 We consider for simplicity an infinitely large
control interval $t\in(-\infty,\infty)$ with natural boundary
conditions $\theta(-\infty)=V(-\infty)=V(+\infty)=0$ and
$\theta(+\infty)=\pi$. This condition implies that the system is excited from and de-excited  to the ground-state, remaining there
 after the interaction with the control field. It is clear  that these are the   boundary conditions compatible with a minimization of $Q_{22}$ (see Fig.~1).
 Eq.~(\ref{sine-gordon})  can be integrated analytically (it is
 mathematically equivalent to the pendulum equation).  The resulting
 optimal field envelope $V_{opt}(t)$ is given by the expression
\begin{equation}
\label{soliton_field}
V_{opt}(t) = {(\sqrt{\lambda} \cosh(t/\sqrt{\lambda/ \mu^2})}^{-1}.
\end{equation}
The Lagrange multiplier $\lambda$ is determined from the normalization
condition $\int_{-\infty}^{+\infty} V^2(t) dt = E_0$ for the pulse energy  and is given
by $\lambda = 4/(\mu E_0)^2$.

Eq.~[\ref{soliton_field}] represents  the well known soliton solution for the pulse shape and is shown in the inset of Fig.~1.
This  result shows that the soliton wave does not only  propagate
  without shape changes\cite{losses}
 but it also minimizes the energy losses,
 which are proportional to $Q_{22}$  in the limit of the weak relaxation and dephasing.
 This fact can be clearly shown in Fig.~1, where we show the
 integrated population $\int_{-\infty}^{t} \rho_{22}(t') dt'$ over the control interval for the optimal pulse shape and for a square pulse having the same area and the same energy as the optimal.
 From the figure it is clear that $Q_{22}$, and therefore the losses, is smaller for the soliton pulse shape.

Using other asymptotic values of the pulse area
$\theta(+\infty)=N\:\pi$, with $N=2,3,...$, one can immediately reproduce
 $2 \pi, 3 \pi,\ldots$ soliton solutions.
 These soliton shapes are the optimal pulses  corresponding to different
values of the pulse energy, which are, of course, higher than the energy of
 the shape of Eq.~[\ref{soliton_field}].

Now we turn to the "standard" optimal control problem at fixed time and
 show how the analytical solution to the
problem of maximization of an objective at a particular time $t_{control}$ arises
naturally as a limiting case of our theory.
 If the quantity to be maximized is, for instance, the population of the upper level at time $t_{control}$, $\rho_{22}(t_{control})$, then the Lagrangian density becomes, with the help of Eq.~(\ref{objective})
\begin{equation}
\label{lagrang_simple1}
{\cal{L}}_{\delta} = {\rho_{22}(t) } \delta(t-t_{control}) +\lambda  \dot{\theta}^2(t)/\mu^2,
\end{equation}
where the delta function $\delta(t-t_{control})$ accounts for the modification of the fitness density.
The optimal pulse shape can only be obtained analytically for
$\gamma_1=\gamma_2=0$, i.e., if  relaxation and dephasing effects are neglected. In this case  the corresponding Euler-Lagrange equation
 reads
\begin{equation}
\label{sine-gordon1}
2\lambda \ddot{\theta}(t)- \mu^2 \delta(t-t_{control}) \sin(2   \theta(t))=0.
\end{equation}
Integrating Eq.~(\ref{sine-gordon1}) one obtains the pulse area as a
linear function $\theta(t)=At+B$. By substituting the boundary
conditions $\theta(0)=0, \theta(t_{control}) = \pi/2$, we find that
the solution of Eq.~(\ref{sine-gordon1}) is a field with a constant
amplitude
\begin{equation}
\label{solution1} V_{opt}(t) = \frac{\pi}{2  \mu \; t_{control} },
\end{equation}
with energy
 $E_0=\pi^2/(4   \mu^2 t_{control})$, measured in the time-interval $[0,t_{control}]$.
 This result reflects the fact that, among all pulses with area
equal to $\pi/2$, that which minimizes the energy has {\it
time-independent shape}. It must be pointed out that this
analytical solution corresponds to the global extremum of the
Lagrangian as long as  the RWA is applicable.

It is important recall that Eq.~(\ref{solution1}) is a new result
and should not be confused with the trivial fact that a $\pi/2$
pulse, when limited to a constant amplitude, produces complete
inversion.

In Fig.~2 we compare the analytical result of Eq.~ [\ref{solution1}]
with a numerical solution obtained by Zhu et al. using
OCT\cite{rabitz_constant}. In Ref.~\cite{rabitz_constant}, the optimal
field to induce population inversion between two levels of the Morse
potential at a particular time was calculated using an iterative
numerical technique to integrate the OCT equations. The obtained field
consists of a single frequency (resonant with the difference of the
level-energies) and a time-dependent amplitude, shown as a dashed line
in Fig.~2. In order to reproduce the same physical situation and use
the same parameters as in Ref.~\cite{rabitz_constant} we calculated
the dipole matrix element $\mu$ as $\mu= \langle\psi_0 |
\hat{\mu}\psi_1\rangle$, where $\hat{\mu}(r)=\mu_0 r
e^{-r/r_0}$,$\mu_0=3.088$, and $r_0=0.6$.  $\psi_0(r)$ and $\psi_1(r)$
correspond to the ground and the first excited state eigenfunctions of
the Morse potential, which is given by $V(r)=D_0 (\exp(-
\beta(r-r^*))-1)^2-D_0$, with $D_0=0.1994$, $\beta=1.189$ and
$r^*=1.821$ \cite{rabitz_constant}.
 Thus, we used Eq.~[\ref{solution1}] to determine the magnitude of
 the constant optimal amplitude.
 Our analytically calculated optimal field, which corresponds
in fact to the true global extremum,
  is shown as a thin solid line in Fig.~2.
 Note that the shape obtained by Zhu et al.  is close to that of the optimal
 field obtained by us. However, it is clear that it does not correspond to the
 global extremum of the problem.  Moreover, the numerically determined shape
 shows a slight asymmetry, which gives rise to a broadening of its
 Fourier spectrum, i.e., to a less effective coupling to the two-level
 system.

 This example shows that our analytical approach can be used to check
 the ability of different numerical methods to avoid local extrema.

 Summarizing, we have obtained analytical solutions for the optimal
 shape of external fields to control populations in two-level systems
 over finite time-intervals and at particular control times. Our
 obtained optimal shapes constitute the global extrema of the control
 problem, in contrast to previous numerical solutions.  Our results
 can be used as a basis to solve optimal control problems in materials
 which are well described by collections of two-level systems.

This work has been supported by the Deutsche Forschungsgemeinschaft
through SFB 450 and by the Spanish MCyT through  BFM2002-03241 and
the Program Ramon y Cajal.


\begin{figure}[h]
\begin{center}
\includegraphics[bb=19 414 436 715, width=0.85\textwidth]{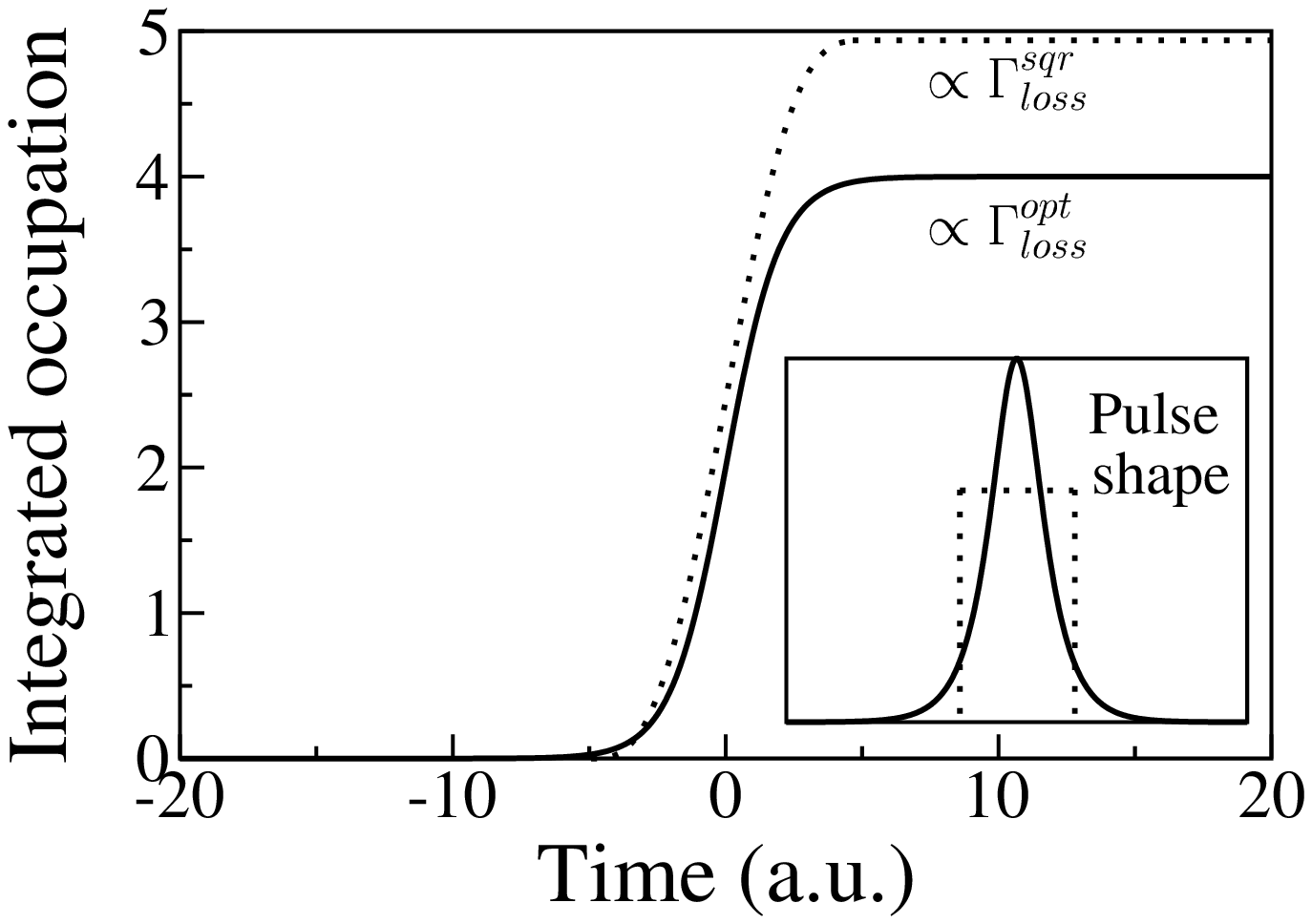}
\caption {\label{soliton}
 Integrated occupation of the upper level $\int_{-\infty}^{t} \rho_{22}(t') dt'$
 for the optimal (soliton) pulse (solid line) and for a square pulse
 (dashed line), shown in the inset figure. Note that $Q_{22}= \int_{-\infty}^{\infty}
 \rho_{22}(t') dt'$
 is proportional to the losses $\Gamma_{loss}$. Inset figure: pulse shape for the optimal control
field $V_{opt}(t)$
 to achieve self-induced
transparency through minimization of the losses in a two-level
system (solid line), and for a square pulse having the same pulse
area and energy (dashed line).}
\end{center}
\end{figure}

\begin{figure}[h]
\begin{center}
\includegraphics[width=0.7\textwidth]{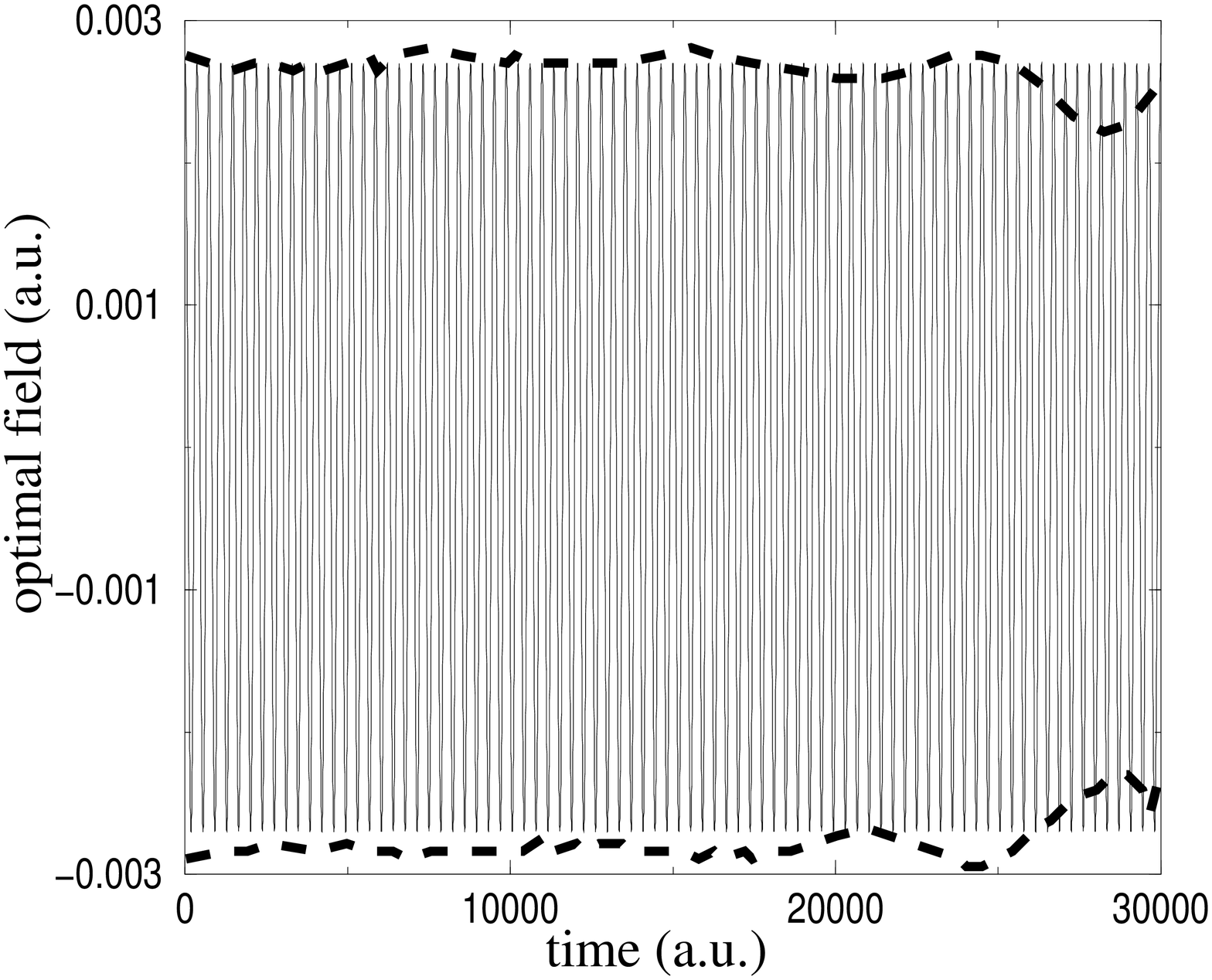}
\vspace*{0.3cm}
\caption {\label{constantfield}
  Thin solid line: analytical solution for the optimal control field
  $V_{opt}(t) \cos{(\omega t)}$, with energy $E_0=\pi^2/(4 \mu^2
  t_{control})$ (see text) to produce inversion of the population at
  $t_{control}=30000$ (a.u.).  The optimal amplitude $V_{opt}(t)$ is
  time-independent [see Eq.~(\ref{solution1})].  Dashed line:
  numerical result $V_{oct}(t)$ for the field amplitude for the
 same problem and the same  system-parameters obtained
 in Ref.~2 using optimal control theory. }
\end{center}
\end{figure}



\begin{thebibliography}{99}
\bibitem{losses} Since we neglect changes of the field along the
 spatial variable one can treat the  Eq.~[\ref{sine-gordon}] as the
sine-Gordon equation in the limit of the optically thin media.
\end{thebibliography}
\end{document}